\documentclass[a4paper]{article}
\usepackage{amsmath}
\usepackage{graphicx}
\usepackage{epstopdf}
\usepackage{cite}
\title{The Inverse Problem for the Dipole Field}
\author{V. Epp\thanks{E-mail: epp@tspu.edu.ru} \,and J. Janz\\
{\small Tomsk State Pedagogical University, pr. Komsomolsky 75, 634041 Tomsk, Russia}
}
\date{}
\begin{document}
\maketitle
\begin{abstract}
The Inverse problem for an electromagnetic field produced by a dipole is solved. It is assumed that the field of an arbitrary changing dipole is known. Obtained formulae allow calculation of the position and dynamics of the dipole which produces the measured field. The derived results can be used in investigations on radiative process in solids caused by changing of the charge distribution. For example, generation of the electromagnetic field caused by oscillations of atoms or electron gas at the trace of a particle channeling in a crystal, or fields arising at solids cracking or dislocation formation -- in any case when one is interested in the details of the dipole field source.
\end{abstract}

\mbox{}

\mbox{}

\vspace{5mm}

There are many applications where the problem of calculation of the source of the known electromagnetic field is arisen. For example, when the first undulators were developed, a question occurred on how to construct the undulator field in order to produce the radiation of given properties. Some ways of solution of this inverse problem was suggested in Refs \cite{nik76,nik77,bes82}. An alternative approach concerning radiation in a ''short magnet'' was suggested by authors of Refs \cite{bagrovNIM,bagrovPhRev}. The general solution of the inverse radiation problem is presented in Ref. \cite{bagrov85}. The listed above papers present the methods of calculation of the particle motion if the produced electromagnetic field is known. The same problem is actual in astrophysics in research of distant sources  of radiation. A general solution of the inverse problem for an arbitrary moving charged particle is given in Ref \cite{epp04}.

Another elementary and widespread source of electromagnetic field is a variable electric or magnetic dipole. Relativistic particle physics presents a great variety of the dipole radiation phenomena. One of the interesting example for inverse problem application is the field generated by atoms and electron gas exited by channeling particles in a crystal \cite{book}. Of spatial interest are oscillations of atoms at a trace of a channeling particle. Solution of inverse problem for a field produced by an arbitrary changing dipole could be a good instrument of investigations in this area. Though the solution of the inverse problem for a dipole field is of great interest in very different areas of physics -- for example, in forecast of earthquakes or investigation of development of cracks in crystals (see a review in Ref. \cite{rew}) -- the complete solution of the inverse problem for the dipole field has not been published until now. The solution in a particular case of the static  dipole field was obtained in Refs (\cite{EppDip,EppKop}).

Electric $\vec{E}(t)$ and magnetic $\vec{H}(t)$ fields of a dipole momentum $\vec{p}(t')$ at a distance $r$ is represented by the formulae:
\begin{eqnarray}\label{K-1}
\vec{E}(t)&=&\frac{(\vec{n}\times(\vec{n}\times\ddot{\vec{p}})))}{rc^2}+\frac{3\vec{n}(\vec{n}\cdot\dot{\vec{p}})-\dot{\vec{p}}}{r^2c}+
\frac{3\vec{n}(\vec{n}\cdot\vec{p})-\vec{p}}{r^3},\\
\label{K-2}
\vec{H}(t)&=&\frac{1}{r^2c}(\dot{\vec{p}}\times\vec{n})+\frac{1}{rc^2}(\ddot{\vec{p}}\times\vec{n}),
\end{eqnarray}
where $\vec{n}=\vec{r}/r$, $\vec{r}$ is the radius-vector from the dipole moment to the observation point, $c$ is the speed of light. Dots denote the derivatives with respect to time, the retarded time $t'$  is connected with time $t$ by relation
\begin{eqnarray}\label{tt}
t=t'+\frac{r}{c}.
\end{eqnarray}
Eq. (\ref{K-2}) shows that vector $\vec{H}$ is changing in a plane orthogonal to vector $\vec{n}$, which allows, basically, to define the vector $\vec{n}$. For this purpose we find the vector production $(\vec{H}\times\dot{\vec{H}})$. Taking derivative from Eq. (\ref{K-2}) with respect to $t$ and having in mind that $dt=dt'$, we obtain:
\begin{eqnarray}\label{K-2-2}
(\vec{H}\times\dot{\vec{H}})=\frac{c}{r^7}\vec{n}(\vec{n}\cdot(\vec{P}'\times\vec{P}{''})),
\end{eqnarray}
where $\vec{P}=\vec{p}\,'+\vec{p}$, stroke denotes the derivative with respect to dimensionless time $\tau=ct/r$. The left-hand side of the last equation gives the direction of angular velocity of vector $\vec{H}$. We see that both vectors $\vec{H}$ and $\vec{P}\,'$ rotate around the vector $\vec{n}$ in the same sense. The vector product $(\vec{H}\times\dot{\vec{H}})$ can be either parallel or antiparallel to the vector $\vec n$ depending on dipole dynamics. We assume further that $(\vec{H}\times\dot{\vec{H}})\neq 0$. Then we have for the unit vector $\vec{n}$:  
\begin{eqnarray}\label{N-1}
\vec{n}=\pm\frac{(\vec{H}\times\dot{\vec{H}})}{|(\vec{H}\times\dot{\vec{H}})|}\, .
\end{eqnarray}
The sign in this equation can be defined from the condition that the average energy flow defined by Poynting vector
\begin{eqnarray}\label{N-2}
\vec{S}=\frac{c}{4\pi}(\vec{E}\times\vec{H})
\end{eqnarray}
is directed from the dipole to the observation point. Indeed, let us find projection of $\vec{S}$ onto vector $\vec{n}$ by use of Eqs (\ref{K-1}) and (\ref{K-2}):
\begin{eqnarray}
(\vec{n}\cdot\vec{S})=\frac{c}{4\pi r^6}\left\{(\vec{n}\times\vec{p}\,{''})^2+(\vec{n}\times\vec{p} \,')\cdot(\vec{n}\times\vec{p}\,{''})+(\vec{n}\times\vec{P})\cdot(\vec{n}\times\vec{P}\, ')\right\}.\nonumber
\end{eqnarray}
The last equation can be written in the form
\begin{eqnarray}\label{N-4}
(\vec{n}\cdot\vec{S})=\frac{c}{4\pi
r^6}\left\{(\vec{n}\times{\vec{p}}\,{''})^2+\frac{1}{2}\frac{d}{d\tau}\left((\vec{n}\times\vec{p}\,')^2+(\vec{n}\times\vec{P})^2\right)\right\}.
\end{eqnarray}
We assume that the dipole moment $p$ varies within a limited interval. Then after avereging over time $\tau$ from $-\infty$ to $\infty$ the value of second term terns to zero and the average rate of the first term represents the well known value of the dipole radiation:
\begin{eqnarray}\label{N-5}
\overline{(\vec{n}\cdot\vec{S})}=\frac{c}{4\pi r^6}
\overline{(\vec{n}\times{\vec{p}}\,{''})^2}.\nonumber
\end{eqnarray}
Thus, the true direction of vector $\vec n$ in Eq. (\ref{N-1}) is defined by condition  $\overline{(\vec{n}\cdot\vec{S})}>0.$ The second term in Eq. (\ref{N-4}) can be both positive and negative, and describes the oscillatory part of the energy transition in the near-field zone. Generally speaking, direction of this part of energy flow is not parallel with the vector $\vec n$. 

Now, knowing the vector $\vec{n},$ we can integrate Eqs (\ref{K-1}) and
(\ref{K-2}) in order to calculate the dipole moment $\vec{p}(t')$.  After first integration of Eq. (\ref{K-2}) we obtain:
\begin{eqnarray}\label{N-6}
\vec{H}_1=\frac{1}{r^3}(\vec{P}\times\vec{n}),
\end{eqnarray}
where
\begin{eqnarray}
\vec{H}_1=\int \vec{H}(t)d\tau\nonumber
\end{eqnarray}
is a function known to an additive constant. We adopt a Cartesian system of coordinates with the axis $OX$ parallel to the vector  $\vec n$. Then Eq. (\ref{N-6}) consist of two scalar equations:
\begin{eqnarray}\label{K-18}
H_{1y}&=&\frac{1}{r^3}(p'_z+p_z),\\
\label{K-19}
H_{1z}&=&-\frac{1}{r^3}(p'_y+p_y).
\end{eqnarray}
The general solution of this system of differential equations is as following:
\begin{eqnarray}\label{K-22}
\vec p_\perp=e^{-\tau}\left(\vec{p}_{0\perp}+r^3\int(\vec{n}\times\vec{H}_1)e^\tau
d\tau\right),
\end{eqnarray}
where $\vec{p}_\perp$ is the projection of vector $\vec{p}$ onto the plane
$YZ$ and  $\vec{p}_{0\perp}$ is an arbitrary vector in the $YZ$ plane.

In order to find the $X$ component of vector $\vec{p}$ we take a scalar production of Eq. (\ref{K-1}) and vector $\vec{n}.$ This gives:
\begin{eqnarray}
(\vec{n}\vec{E})=\frac{2}{r^3}(p'_x+p_x).\nonumber
\end{eqnarray}
Hence:
\begin{eqnarray}\label{N-7}
p_x=e^{-\tau} \Big(p_{0x}+\frac{r^3}{2}\int (\vec{n}\vec{E})e^\tau
d\tau\Big),
\end{eqnarray}
where $p_{0x}$ is an arbitrary constant.
Combining Eqs (\ref{K-22}) and (\ref{N-7}) we obtain the general solution of Eqs (\ref{K-1}) and (\ref{K-2}):
\begin{eqnarray}\label{N-8}
\vec{p}(t')=e^{-\tau}\left\{\vec{p}_{0}+r^3\int
\big[(\vec{n}\times\vec{H}_1)+\frac{1}{2}\vec{n}(\vec{n}\cdot\vec{E})\big]e^\tau
d\tau\right\}.
\end{eqnarray}
Here $\vec{p}_0=\vec{p}_{0\perp}+\vec{n}p_{0x}$ is an arbitrary vector. Up to now we have not calculated the distance $r$ between the observation point and dipole. Hence we do not know the relation between times $\tau$ in the right-hand side of Eq. (\ref{N-8}) and $t'$ in its left-hand side.

Substituting the received solution (\ref{N-8}) in initial Eqs (\ref{K-1}) and (\ref{K-2}) we see that Eq. (\ref{K-2})
is satisfied identically but Eq. (\ref{K-1}) gives:
\begin{eqnarray}\label{N-9}
\vec{E}=\vec{n}(\vec{n}\cdot\vec{E})-(\vec{n}\times\vec{H})-\frac{1}{r^3}
e^{-\tau}\vec{p}_{0\perp}-e^{-\tau}\int(\vec{n}\times\vec{H}_1)e^{\tau}d\tau.
\end{eqnarray}
The last relation shows that the vectors $\vec{E}$ and $\vec{H}$ are not independent -- they are bound by Eq. (\ref{N-9}). The relation between electric and magnetic fields of dipole can be found easily from Eqs (\ref{K-1}) and (\ref{K-2}) by inspection. To do this we take a vector product of the vector $\vec{n}$ with Eq. (\ref{K-1}) and extract Eq. (\ref{K-2}). As a result we get:
\begin{eqnarray}\label{N-101}
\vec{H}-(\vec{n}\times\vec{E})=\frac{1}{r^3}(\vec{n}\times\vec{p}).
\end{eqnarray}
Substituting $(\vec{n}\times\vec{p})$ from this equation into Eq. (\ref{K-2}), we find the desired coupling equation:
\begin{eqnarray}\label{N-10}
\frac{r^2}{c^2}\ddot{\vec{H}}+\frac{r}{c}\dot{\vec{H}}+\vec{H}=\frac{r^2}{c}(\vec{n}\times\ddot{\vec{E}})+\frac{r}{c}(\vec{n}\times\dot{\vec E}).
\end{eqnarray}
It is easy to show that Eq. (\ref{N-9}) can be transformed to Eq. (\ref{N-10}) by multiplying into $e^\tau$ and taking the second derivative with respect to $\tau$. Thus, the solution (\ref{N-8}) satisfies the  initial Eqs (\ref{K-1}) and (\ref{K-2}).

The distance between the dipole and observation point can be calculated by use of Eq. (\ref{N-10}). It is a quadratic equation with respect to $r$:
\begin{eqnarray}\label{N-11}
\vec A \rho^2 +\vec B \rho +\vec C=0,
\end{eqnarray}
where vectors $\vec A=\ddot{\vec H}-(\vec n\times\ddot{\vec E})$, $\vec B=\dot{\vec H}-(\vec n\times\dot{\vec E})$, $\vec C=\vec H$ are known, and $\rho=r/c$.
The vectors $\vec A$, $\vec B$ and $\vec C$ lie in a plane which is orthogonal to vector $\vec n$. Hence, Eq. (\ref{N-11}) is equal to the system of two scalar equations. Compatibility of the system requires certain relation between the coefficients of the equations which can be written in the form  $(\vec A\times\vec C)^2=((\vec A\times\vec B)\cdot(\vec B\times\vec C))$. Then Eq. (\ref{N-11}) has the solution
\begin{eqnarray}\label{N-12}
r=c\,\frac{(\vec n\cdot(\vec C\times\vec A))}{(\vec n\cdot(\vec A\times\vec B))}=c\,\frac{(\vec n\cdot(\vec B\times\vec C))}{(\vec n\cdot(\vec C\times\vec A))}.
\end{eqnarray}
In order to verify the obtained solution we represent the vectors  $\vec A$, $\vec B$ and $\vec C$ in terms of dipole moment. This can be made by use of Eq. (\ref{N-101}): $\vec A=(\ddot{\vec p}\times\vec n)/r^3$, $\vec B=(\dot{\vec p}\times\vec n)/r^3$. The vector products take the form:
$$(\vec A\times\vec B)=\frac{\vec n}{r^6}(\vec n\cdot(\ddot{\vec p}\times\dot{\vec p})),\, (\vec C\times\vec A)=\frac{\vec n}{cr^5}(\vec n\cdot(\ddot{\vec p}\times\dot{\vec p})), \, (\vec B\times\vec C)=\frac{\vec n}{c^2r^4}(\vec n\cdot(\ddot{\vec p}\times\dot{\vec p})).$$
The last relations show that the obtained solution for $r$ is a positive constant, though the nominators and denominators in Eq. (\ref{N-12}) are time-dependent. But the expressions for $r$ are not valid if $(\vec n\cdot(\ddot{\vec p}\times\dot{\vec p}))=0$, i.e. if vector $\dot{\vec p}$ is varying in a plane going through vector $\vec n$. In this case vectors $\vec P\,'$ è $\vec P\,''$ are varying in the same plane, hence, the vector product in Eq. (\ref{K-2}) are equal to zero. As it was mentioned above this case is not under consideration in the current paper. 
Thus, the solution of the stated problem is given by Eqs (\ref{N-1}), (\ref{N-8}) and (\ref{N-12}) if $(\vec H\times\dot{\vec H})\neq 0$. And even if $(\vec H\times\dot{\vec H})=0$, Eq. (\ref{N-8}) for $\vec{p}(t')$ still remains valid.

As an example we consider an electromagnetic field of oscillating dipole, say the field of an ion oscillating at specific site in the lattice of a crystal. Let the electric dipole have the time dependence
\begin{eqnarray}
p_x=p_{0x}\cos \omega t, \quad p_y=p_{0y}\sin \omega t, \quad p_z=p_{0z},
\end{eqnarray}
where $p_{0x}, \, p_{0y}, \, p_{0z}, \, \omega$ are some given constants. We have calculated the electric and magnetic fields $\vec E(t)$ and $\vec H(t)$ at a distance $\vec r$ from the dipole and have saved the results as a numerical functions of $t$. Then we have used the saved data for calculation of $\vec p(t)$ and $\vec r(t)$ according to Eqs (\ref{N-1}), (\ref{N-8}) and (\ref{N-12}). The result coincides with the initial functions with accuracy up to $10^{-8}$. The error occurs due to the used numerical method of calculation. The distance $r$ due to the errors becomes a random function of time as shown in Fig. 1 for initial distance $r\omega/c=100$.
\begin{figure}[h]
 \includegraphics{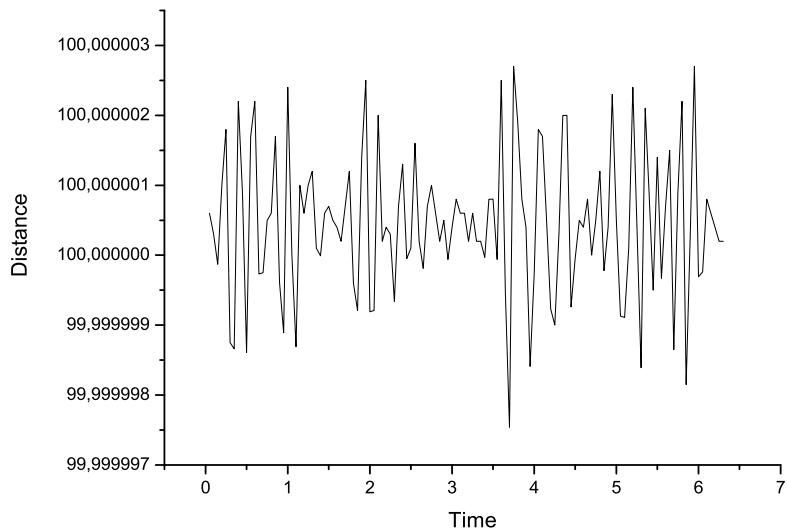}
    \caption{The reconstructed distance $r$ in terms of $c/\omega$ as a function of $Time=\omega t$}
\end{figure}
\section*{Acknowledgment}
The work of V. Epp has been supported by grant for LRSS, project No 4489.2006.2.

\end{document}